\documentclass[a4paper,12pt, epsfig]{article}
\usepackage{epsfig}
\usepackage{amssymb}
\usepackage{amsfonts}
\usepackage{graphics}

\newskip\humongous \humongous=0pt plus 1000pt minus 1000pt

\newif\ifdtup

\catcode`\@=11

\@addtoreset{equation}{section}
\def\theequation{\thesection.\arabic{equation}}

\def\@normalsize{\@setsize\normalsize{15pt}\xiipt\@xiipt
\abovedisplayskip 14pt plus3pt minus3pt%
\belowdisplayskip \abovedisplayskip
\abovedisplayshortskip \z@ plus3pt%
\belowdisplayshortskip 7pt plus3.5pt minus0pt}

\def\small{\@setsize\small{13.6pt}\xipt\@xipt
\abovedisplayskip 13pt plus3pt minus3pt%
\belowdisplayskip \abovedisplayskip
\abovedisplayshortskip \z@ plus3pt%
\belowdisplayshortskip 7pt plus3.5pt minus0pt
\def\@listi{\parsep 4.5pt plus 2pt minus 1pt
      \itemsep \parsep
      \topsep 9pt plus 3pt minus 3pt}}

\relax

\catcode`@=12

\catcode`\@=11

\def\section{\@startsection{section}{1}{\z@}{3.5ex plus 1ex minus
    .2ex}{2.3ex plus .2ex}{\large\bf}}

\def\thesection{\arabic{section}}
\def\thesubsection{\arabic{section}.\arabic{subsection}}

\def\appendix{\setcounter{section}{0}
  \def\thesection{Appendix \Alph{section}}
  \def\thesubsection{\Alph{section}.\arabic{subsection}}
  \def\theequation{\Alph{section}.\arabic{equation}}}

\def\SymBoxes#1#2#3#4{\newdimen\un@t \un@t#3%
\raisebox{#1}{\rule{#2\un@t}{#4}\hskip-#2\un@t
\@tempdimb\un@t \advance\@tempdimb by-#4\@tempcntb#2\relax%
\@whilenum{\@tempcntb>0}\do{
\rule{#4}{\un@t}\hskip\@tempdimb \advance\@tempcntb by\m@ne}%
\hskip-#2\un@t \rule[\un@t]{#2\un@t}{#4}%
\rule[\un@t]{#4}{#4}\hskip-#4
\rule{#4}{\un@t}}\hskip-#4}                

\begin{document}


\newcommand{\dd}{\textrm{d}}

\newcommand{\beq}{\begin{equation}}
\newcommand{\eeq}{\end{equation}}
\newcommand{\bea}{\begin{eqnarray}}
\newcommand{\eea}{\end{eqnarray}}
\newcommand{\beas}{\begin{eqnarray*}}
\newcommand{\eeas}{\end{eqnarray*}}
\newcommand{\defi}{\stackrel{\rm def}{=}}
\newcommand{\non}{\nonumber}
\newcommand{\bquo}{\begin{quote}}
\newcommand{\enqu}{\end{quote}}
\renewcommand{\(}{\begin{equation}}
\renewcommand{\)}{\end{equation}}
\def\de{\partial}
\def\Om{\ensuremath{\Omega}}
\def\Tr{ \hbox{\rm Tr}}
\def\H{ \hbox{\rm H}}
\def\HE{ \hbox{$\rm H^{even}$}}
\def\HO{ \hbox{$\rm H^{odd}$}}
\def\HEO{ \hbox{$\rm H^{even/odd}$}}
\def\HOE{ \hbox{$\rm H^{odd/even}$}}
\def\HHO{ \hbox{$\rm H_H^{odd}$}}
\def\HHEO{ \hbox{$\rm H_H^{even/odd}$}}
\def\HHOE{ \hbox{$\rm H_H^{odd/even}$}}
\def\K{ \hbox{\rm K}}
\def\Im{ \hbox{\rm Im}}
\def\Ker{ \hbox{\rm Ker}}
\def\const{\hbox {\rm const.}}
\def\o{\over}
\def\im{\hbox{\rm Im}}
\def\re{\hbox{\rm Re}}
\def\bra{\langle}\def\ket{\rangle}
\def\Arg{\hbox {\rm Arg}}
\def\exo{\hbox {\rm exp}}
\def\diag{\hbox{\rm diag}}
\def\longvert{{\rule[-2mm]{0.1mm}{7mm}}\,}
\def\a{\alpha}
\def\dag{{}^{\dagger}}
\def\tq{{\widetilde q}}
\def\p{{}^{\prime}}
\def\W{W}
\def\N{{\cal N}}
\def\hsp{,\hspace{.7cm}}
\def\bo{\ensuremath{\hat{b}_1}}
\def\bfo{\ensuremath{\hat{b}_4}}
\def\co{\ensuremath{\hat{c}_1}}
\def\cfo{\ensuremath{\hat{c}_4}}
\newcommand{\C}{\ensuremath{\mathbb C}}
\newcommand{\Z}{\ensuremath{\mathbb Z}}
\newcommand{\R}{\ensuremath{\mathbb R}}
\newcommand{\rp}{\ensuremath{\mathbb {RP}}}
\newcommand{\cp}{\ensuremath{\mathbb {CP}}}
\newcommand{\vac}{\ensuremath{|0\rangle}}
\newcommand{\vact}{\ensuremath{|00\rangle}                    }
\newcommand{\oc}{\ensuremath{\overline{c}}}

\newcommand{\Vol}{\textrm{Vol}}

\newcommand{\half}{\frac{1}{2}}

\begin{titlepage}
\begin{flushright}
SISSA-63/2008/EP
\end{flushright}
\bigskip
\def\thefootnote{\fnsymbol{footnote}}

\begin{center}
{\large {\bf
Which BPS Baryons Minimize Volume?
  } }
\end{center}

\begin{center}

{\large Jarah Evslin\footnote{ evslin@sissa.it}}

\end{center}

\begin{center}
\textit{{\it Scuola Internazionale Superiore di Studi Avanzati (SISSA),\\
Strada Costiera, Via Beirut n.2-4, 34013 Trieste, Italia}}
\end{center}

\begin{center}

{\large Stanislav Kuperstein\footnote{skuperst@ulb.ac.be}}

\end{center}

\begin{center}
\textit{{\it 
Theoretische Natuurkunde, Vrije Universiteit Brussel and\\The International Solvay Institutes\\
Pleinlaan 2, B-1050 Brussels, Belgium \\
}}
\end{center} \vfil

\renewcommand{\thefootnote}{\arabic{footnote}}

\noindent
\begin{center} {\bf Abstract} \end{center}

\noindent

A BPS 3-cycle in a Sasaki-Einstein 5-manifold in general does not minimize volume in its homology class, 
as we illustrate with several examples of non-minimal volume BPS cycles on the 5-manifolds $Y^{p,q}$.  
Instead they minimize the energy of a wrapping D-brane, extremizing a generalized calibration.
We present this generalized calibration and demonstrate that it reproduces both the Born-Infeld and the Wess-Zumino parts of the D3-brane energy.

\vfill

\begin{flushleft}
{\today}
\end{flushleft}
\end{titlepage}

\hfill{}


\setcounter{footnote}{0}

\section{Introduction}

A D-brane wrapping a BPS cycle minimizes the energy in its (twisted) homology class \cite{Harvey:1982xk, gcalpapers1}.  
In the absence of fluxes, the energy is equal to the volume of the wrapped cycle and so BPS 
D-branes wrap minimal volume cycles \cite{Harvey:1982xk}. 
In the presence of fluxes, the energy is equal to the sum of a Born-Infeld contribution, 
which contains the volume, and Wess-Zumino contributions, which consist of various fluxes.  
Therefore in general energy and volume are not minimized by the same cycles \cite{gcalpapers1,gcalpapers2}.

In this note we will understand this observation from the viewpoint of generalized calibrations.  In particular we will present a generalized calibration that calibrates \mbox{3-cycles} on all Sasaki-Einstein 5-manifolds.  While generalized calibrations have been studied extensively in the context of generalized Calabi-Yau 6-manifolds, to our knowledge, 
the only Sasaki-Einstein manifold on which a generalized 
calibration has been presented so far is the 5-sphere \cite{HackettJones:2004yi}.
We will report then a number of explicit examples of non-BPS 3-cycles with lower volumes than 
BPS 3-cycles in the same homology class on Sasaki-Einstein manifolds $Y^{p,q}$ (see \cite{Ypq}).

A generalized calibration on a manifold $M$ endowed with a closed 
$(p+2)$-form $F_{p+2}$ is a $p$-form $\phi_p$ such that when pulled back to any $p$-dimensional subbundle 
$E$ of the tangent bundle $TM$:
\beq
i:E\hookrightarrow TM\hsp i^*\phi_p \leqslant \dd \Vol_E \label{ineq},
\eeq
where $\dd \Vol_E$ is the volume form on the subbundle, and such that there exists some unit 
Killing vector $\xi$ satisfying:
\beq
\dd \phi_p =i_\xi F_{p+2}. 
\label{ig}
\eeq
If we compactify type II string theory on a $(p+2)$-dimensional manifold $M^{p+2}$, 
with $F_{p+2}$ the RR $(p+2)$-form field strength, then the Noether energy density of a D$p$-brane with respect to the light-like Killing vector $\xi$ is the sum of the 
Born-Infeld contribution $\Vol_E$ and the Wess-Zumino contribution $i_\xi C_{p+1}$, 
where $C_{p+1}$ is the potential for $F_{p+2}$ in some gauge\footnote{See Section \ref{townsec} 
for a more detailed discussion of the gauge choice.}.  
The brane has minimal energy in its homology class when the inequality in (\ref{ineq}) 
is saturated \cite{gcalpapers1,gcalpapers2}.  In this case the cycle $\Sigma^p$ wrapped by the 
D$p$-brane is said to be calibrated by $\phi_p$.  In particular, the D$p$-brane can only be 
BPS with respect to the Killing spinor $\epsilon$ if the cycle is calibrated with respect to the Killing vector:
\beq
\xi_\mu=\overline\epsilon\Gamma_\mu\epsilon, 
\label{xi}
\eeq
where $\Gamma_\mu$'s are gamma matrices.

In Section \ref{calsec} we will introduce a generalized calibration for 
Sasaki-Einstein 5-folds with a specific RR flux.  In Section 
\ref{Stansec} we will explicitly calculate the volumes of some BPS \mbox{3-cycles} in $Y^{p,q}$ 
and we will find that sometimes non-BPS cycles in the same homology class have smaller volumes than BPS cycles.  
In Section \ref{townsec} we specialize the argument of \cite{gcalpapers1} 
that D3-branes wrapping BPS 3-cycles of unequal volumes nonetheless have the same energy to the case of 
generalized calibrations on Sasaki-Einstein 5-manifolds.  This argument is then applied to BPS \mbox{3-cycles} in $Y^{p,q}$. 
Finally in the conclusion we provide a formula for the energies of branes wrapping non-BPS cycles.  
The generalized calibration implies that these are necessarily greater than those of branes wrapping 
BPS cycles in the same homology class, but we provide an explicit example in which this is indeed the case, 
even though BPS cycles have a larger volume.

\section{The generalized calibration} \label{calsec}

\subsection{The proposal}

To define a generalized calibration one needs to choose the vector $\xi$ in (\ref{ig}).  
The manifold $M^5$ is Euclidean, and so there are no available light or time-like vectors, 
instead we will choose the Reeb vector. Therefore $\xi$ will not satisfy (\ref{xi}).  
Nevertheless the generalized calibration constructed from $\xi$ will summarize the BPS condition, 
because $\xi$ is the spatial part of a light-like Killing vector whose temporal part does not contribute 
to the energy of D-branes wrapping nontrivial cycles in $M^5$
(see Section \ref{townsec} for a related discussion).

More precisely, consider the geometry $AdS_5 \times M^5$ where $C(M^5)$ is the Calabi-Yau cone over 
the $5d$ Sasaki-Einstein space $M^5$.  The preserved $10d$ Killing spinor $\widetilde{\epsilon}$ may be decomposed 
into the tensor product of 
a Killing spinor $\chi$ on $AdS_5$ and a Killing spinor $\epsilon$ on $M^5$:
\beq
\widetilde{\epsilon}=\chi \otimes \epsilon.
\eeq
One may use this Killing spinor to define a one-form:
\beq
\widetilde{\eta}=(\overline{\widetilde{\epsilon}} \Gamma_\mu\widetilde{\epsilon}) \, \dd x^\mu.
\eeq
The temporal part of $\widetilde{\eta}$ comes entirely from the temporal part of the geometry, in the $AdS_5$ factor. 
It will contribute to (\ref{ig}), however $i_{\partial/\partial t}F_5$ pulled back to cycles on the 
Sasaki-Einstein will vanish and so we will not be interested in this contribution.  
Instead, we will be interested only in the contribution $\eta$ of the Sasaki-Einstein part of the Killing spinor:
\beq
\eta=(\overline{\epsilon}\Gamma_\mu\epsilon) \dd x^\mu,
\eeq
which is the contact form of $M^5$ \cite{contact}.  The contact form is dual to the Reeb vector $\xi$. 
Therefore a D3-brane on $M^5$ will be BPS if and only if it saturates the bound (\ref{ineq}) where $\phi_3$ 
satisfies (\ref{ig}) with $\xi$ equal to the unit Reeb vector.

The metric on $M^5$ may be decomposed as:
\beq
\dd s^2_{M^5}= \dd s^2_{KE} + \eta \otimes \eta,
\label{factors}
\eeq
where $ds^2_{KE}$ is K\"ahler-Einstein metric on the subspace of the tangent bundle of $M^5$ 
which is orthogonal to the Reeb vector field.  When $M^5$ is regular, as in the case $M^5=T^{1,1}$, 
then $M^5$ is just a circle fibration over a K\"ahler Einstein base and $\eta$ is the vertical form plus 
connection.
In this section we will 
not restrict our analysis to the $Y^{p,q}$ \cite{Ypq}  or the $L^{a,b,c}$ \cite{Labc} 
family of spaces, considering instead
a general non-singular 5-dimensional Sasaki-Einstein manifold.

The metric of the Calabi-Yau cone $C(M^5)$ over $M^5$ is simply:
\beq
\dd s^2_{C(M^5)}=\dd r^2 + r^2 \dd s^2_{M^5},
\label{ConicMetric:eq}
\eeq
where $r$ is the radial direction on the cone, $M^5$ is embedded at $r=1$.  
If $J_{KE}$ is the K\"ahler form of the 4-dimensional K\"ahler-Einstein metric, then the K\"ahler 
form $J$ of the Calabi-Yau is:
\beq
J = \frac{1}{2} \dd \left( r^2 \eta \right) = r^2 J_{KE}+ r \dd r \wedge \eta,
\eeq
where we used the fact that $\dd \eta = 2 J_{KE}$.
The Calabi-Yau is calibrated by an ordinary (closed) calibration:
\beq
\frac{1}{2} J\wedge J =  r^3 \dd r\wedge J_{KE} \wedge \eta + \frac{1}{2}  r^4 J_{KE} \wedge J_{KE} 
                       = r^3 \dd r \wedge \alpha_3 + r^4 \beta_4,
\eeq
where we have defined:
\beq
\alpha_3 = J_{KE}\wedge\eta \qquad \textrm{and} \qquad \beta_4 = \frac{1}{2} J_{KE}\wedge J_{KE}.
\eeq

The 4-dimensional calibrated cycles $C(\Sigma^3_k)$ of $C(M^5)$ are cones over the 3-cycles $\Sigma^3_k$ of $M^5$.  
This means that the pullbacks of the calibration $\frac{1}{2} J\wedge J$ to the cycles $C(\Sigma^3_k)$ 
are equal to their volume forms:
\beq
I_k : C(\Sigma^3_k) \hookrightarrow C(M^5) \hsp I_k^* \left( \frac{1}{2} J\wedge J \right)=
                                  \dd \Vol_{C(\Sigma^3_k)}= r^3 \dd r\wedge \dd \Vol_{\Sigma^3_k}, 
\label{4i}
\eeq
where $\dd \Vol_{C(\Sigma^3_k)}$ and $\dd \Vol_{\Sigma^3_k}$ are the volume forms on 
$C(\Sigma^3_k)$ and $\Sigma^3_k$ respectively, 
and we have used the conic structure of the $6d$  metric to find the relation between the two volume forms.
Pushing forward via the projection map:
\beq
\pi:C(M^5)\longrightarrow M^5,
\eeq
which integrates away the $\dd r$ factors, we arrive at:
\beq
i_k:\Sigma^3_k \hookrightarrow M^5 \hsp i_k^*(\alpha_3) = \dd \Vol_{\Sigma^3_k},
\eeq
where the embedding $i_k$ is the restriction of the embedding $I_k$ in (\ref{4i}) to $\Sigma^3_k$.  
Therefore $\alpha_3$ pulled back to $M^5$, which we also denote $\alpha_3$, is a 3-form such that, when 
pulled back to the base $\Sigma^3_k$ of a BPS cycle $C(\Sigma^3_k)$, it is equal to the volume form.  This motivates 
the following proposal \cite{Sparks}:

\noindent
\textit{The 3-form  $\alpha_3 = J_{KE} \wedge \eta$ is a generalized calibration for Sasaki-Einstein 5-folds 
with respect to the Reeb vector $\xi$, when $F_5$ from (\ref{ig}) is equal to four times the volume form.}

Notice that in type IIB supergravity compactifications on $AdS_5 \times M^5$, the RR flux 5-form is indeed four times the 
volume form of the Sasaki-Einstein  space $M^5$.  For concreteness we have restricted our attention to Sasaki-Einstein 5-folds, but $\alpha_{2k+1}=\frac{1}{k!}J_{KE}^k\wedge\eta$ is a generalized calibration for any Sasaki-Einstein $(2k+3)$-manifold.

\subsection{The demonstration}

To check this proposal, one must verify that the inequality (\ref{ineq}) holds for all 3-cycles 
$\Sigma^3$ in $M^5$ and also that (\ref{ig}) is satisfied.  Let $\Sigma^3$ be a 3-cycle in $M^5$ 
such that (\ref{ineq}) is not satisfied.  In other words:
\beq
i:\Sigma^3 \hookrightarrow M^5\hsp i^*\alpha_3 > \dd \Vol_{\Sigma^3},
\label{nproj}
\eeq
where $\dd \Vol_{\Sigma^3}$ is the volume form of $\Sigma^3$.  

Let $C(\Sigma^3)$ be the cone over $\Sigma^3$, which has volume form $r^3 \dd r\wedge \dd \Vol_{\Sigma^3}$. 
Multiplying (\ref{nproj}) by $r^3 \dd r$ one finds that the volume form of $C(\Sigma^3)$ is less then the integral of the pullback of a particular four-form $\gamma_4 \equiv r^3 \dd r\wedge\alpha_3$:
\beq
I:C(\Sigma^3) \hookrightarrow C(M^5)\hsp  I^*\gamma_4 = I^* \left( r^3\dd r\wedge\alpha_3 \right) > 
   r^3 \dd r \wedge \dd \Vol_{\Sigma^3} = \dd\Vol_{C(\Sigma^3)}.
\eeq
As the cone $C(\Sigma^3)$ is the product of the radial direction and an orthogonal 3-fold, the pullback to $C(\Sigma^3)$ of a 4-form with all legs along the base is zero.  In particular, the pullback of $\beta_4$ to $C(\Sigma^3)$ is zero, and so the pullback of the calibrating 4-form $\half J\wedge J$ is equal to that of $r^3 \dd r\wedge\alpha_3$.  In summary:
\beq
I^* \left( \half J\wedge J \right) = I^*(r^3 \dd r\wedge\alpha_3) > \dd \Vol_{C(\Sigma^3)}.
\eeq
This is in contradiction with the fact that $\half J\wedge J$ is a calibration on $C(M^5)$, therefore no such 3-cycle $\Sigma^3$ may exist and so $\alpha_3$ satisfies the inequality (\ref{ineq}) with respect to all 3-cycles.

Now we need to show that $\alpha_3$ also satisfies the condition (\ref{ig}). 
Indeed, since \mbox{$\dd \eta=2J_{KE}$} we find that:
\beq
\dd \alpha_3 = 2 J_{KE}\wedge J_{KE}.
\eeq
On the other hand, the volume 6-form of $C(M^5)$ is $\dd \Vol_{C(M^5)} = \frac{1}{6} J \wedge J \wedge J$ and thus the
volume 5-form on $M^5$ is $\dd \Vol_{M^5} = \half J_{KE} \wedge J_{KE} \wedge \eta$. Since the contact form $\eta$ and the Reeb
vector $\xi$ are dual, namely $i_\xi \eta=1$, we finally obtain that:
\beq
\dd \alpha_3 = 4 \cdot i_\xi \dd \Vol_{M^5} = i_\xi F_5
\eeq
in accordance with the condition (\ref{ig}).  Therefore $\alpha_3$ is indeed a generalized calibration for $M^5$ 
with $F_5= 4 \cdot \dd \Vol_{M^5}$.

\section{Calculating volumes of submanifolds} \label{Stansec}

Consider a cone $C(Y^{p,q})$ over a Sasaki-Einstein base $Y^{p,q}$. The cone is Calabi-Yau and so it is 
calibrated by the 4-form $\half J\wedge J$, where $J$ is its K\"ahler form.  The cone $C(Y^{p,q})$ is the K\"ahler quotient 
of $\C^4 \backslash \{ \C^2 \cup \C^2\}$ by a $\C^*$ action under which the complex coordinates $(z_1,z_2,z_3,z_4)$ 
transform with weights $(-p,-p,p-q,p+q)$.  The 4-dimensional submanifolds on which $z_i$ 
vanish are divisors $C(\Sigma_i^3)$ of $C(Y^{p,q})$.  
They are interesting because they are calibrated.  Their volumes are infinite, 
however their volume density is equal to the pullback of $\half J\wedge J$ to their world-volumes \cite{Ypq}.  

At large $N$, we are not interested precisely in the divisors $C(\Sigma_i^3)$ , but rather in their 
3-dimensional bases $\Sigma_i^3$.  
Branes that wrap these bases are also BPS. The near-horizon geometry of a 
stack of $N$ D3-branes at the tip of the conifold $C(T^{1,1})$ 
is $AdS_5\times T^{1,1}$ with $N$ units of RR 5-form flux on the $T^{1,1}$.
Based on the ideas of \cite{WittenBranesBaryonsAds}
it was conjectured in \cite{GubserKlebanov} (see also \cite{HBKandB})
that the BPS (di-)baryons in the dual CFT \cite{KW} are dual to 
D3-branes wrapped on the 3-cycles $\Sigma_i^3$ on the $T^{1,1}$ which are the bases of the divisors $z_i=0$.  
The conjecture relies on the fact that $T^{1,1}$ has the topology of $S^3\times S^2$
and in particular
\cite{KW,CdO,EK1}:
\beq
\H_3 \left( T^{1,1} \right)=\Z
\eeq
and so the homology class of a 3-cycle is a single integer. 
Remarkably, $Y^{p,q}$ (and $L^{a,b,c}$) has the same $S^3\times S^2$ topology and so the conjecture of \cite{GubserKlebanov}
has been naturally extended to the CFT models based on the $AdS_5 \times Y^{p,q}$ geometries
\cite{Ypq}.
The homology classes of the bases of the divisors are just equal 
to the weights of the $\C^*$ quotient, in other words it is $(-p)$ for the base $\Sigma_{1,2}^3$ of the $z_{1,2}=0$ 
cycles and 
$(p-q)$ and $(p+q)$ for the 
$z_3=0$ and $z_4=0$ cycles $\Sigma_{3}^3$ and $\Sigma_{4}^3$ respectively\footnote{ \label{Lens}
It follows from the observation that for $z_4=0$ (and similarly for the other $z_i$'s) 
the D-term condition of the K\"ahler quotient implies that away from the tip 
$z_3 \neq 0$, so we can safely put $\Im(z_3)=0$.
By means of the Hopf map the remaining coordinates $z_1$ and $z_2$ define a cone over $S^2$,
which then has to be quotiented by the residual discrete symmetry $\mathbb{Z}_{p+q}$. The resulting 
space is a cone over the lens space $L(p+q;1)$.
For $z_{1,2}=0$ and $z_3=0$ one finds instead the cones over $L(p;p-1)$ and $L(p-q;1)$ respectively.
Obviously this reproduces the aforementioned homology classes. 
See \cite{Ypq} for more details.
}. 

The goal of this section is to find the minimal volumes of three-spheres representing the third homology class $1\in\Z$
in $Y^{p,q}$ spaces for $q=1$ and arbitrary $p$.  
We begin with a partial review of the results of \cite{EK2},
where the $Y^{p,q}$ spaces were trivialized for arbitrary $p$ and $q=1$ or $2$, restricting our attention to the $q=1$ case.

First we must properly normalize the K\"ahler quotient coordinates $z_i$
discussed in the previous section.
For $q=1$ the D-term condition on the reduction reads:
\beq
p \left(  \vert z_1 \vert^2 + \vert z_2 \vert^2 \right)
- (p-1) \vert z_3 \vert^2 - (p+1) \vert z_4 \vert^2 =0.
\eeq
We are interested only in the $5d$ $Y^{p,1}$ base of the $6d$ $C(Y^{p,1})$ cone. Away from the tip
(where all $z_i$'s vanish)
we can introduce new variables:
\beq
\label{eq:uuvvzzzz}
\left( u_{1},u_{2},v_{1},v_{2} \right) = \Lambda^{-p}
\left(z_{1},z_{2},\sqrt{1- \frac{1}{p}} \bar{z}_{3}, \sqrt{1+\frac{1}{p}}\bar{z}_{4} \right).
\eeq
The normalization factor $\Lambda$ is fixed\footnote
{See \cite{EK2} for a discussion of 
the normalization.} by the requirement that both vectors,
$u$ and $v$, have unit length.
Unlike the conifold case here $\Lambda$ depends not only on the radial coordinate $r$ appearing in the conic metric
(\ref{ConicMetric:eq}) but also on one of the coordinates of the $Y^{p,q}$ base (see below).

Next we notice that under the $U(1)$ gauge transformation of the K\"ahler quotient $u$ and $v$
transform like
\beq
\label{U(1):eq}
\left( u_1, u_2 \right) \to e^{i \lambda p} \left( u_1, u_2 \right)
\qquad \textrm{and} \qquad
\left( v_1, v_2 \right) \to  \left( e^{i \lambda (p-1)} v_1, e^{i \lambda (p+1)} v_2 \right),
\eeq
so the vector $w=(w_1,w_2)$ defined by:
\beq
\label{uvw:eq}
\left(
\begin{array}{c}
w_1 \\ w_2
\end{array}
\right)
=
\left(
\begin{array}{cc}
u_1 & -u_2^\star \\
u_2 &  u_1^\star 
\end{array}
\right)
\left(
\begin{array}{c}
v_1^\star \\ -v_2
\end{array}
\right)
\eeq
transforms like $w \to e^{i \lambda} w$. It also has unit length.
By means of the Hopf fibration $w$ describes an $S^2$. To parameterize  
the remaining $S^3$ we need:
\beq
\widehat{w} = 
  c_{\widehat{w}} \left( w_1^p , w_2^p \right) \qquad \textrm{where} \qquad c_{\widehat{w}}=1/\sqrt{|w_1|^{2p}+|w_2|^{2p}},
\eeq
so the length-one $\widehat{w}$ transforms exactly like $u$:
\beq
\widehat{w} \to e^{i \lambda p} \widehat{w}.
\eeq
With $u$ and $\widehat{w}$ in hand we define a special unitary matrix $X \in SU(2)$:
\beq
X = u \widehat{w}^\dagger - \epsilon u^\star \widehat{w}^\textrm{\small{T}} \epsilon,
\eeq
which is $U(1)$-invariant and thus properly defines an $S^3$. 
To summarize, starting from a $Y^{p,1}$ given by $u$ and $v$, we may find the $w$ and then $X$ that 
describe the $S^2$ and the $S^3$ respectively. Alternatively beginning with $X$ and $w$ we can determine $\widehat{w}$
from $w$ and then $u$ from the identity:
\beq
u = X \widehat{w},
\eeq
that follows directly from the definition of $X$. Finally, (\ref{uvw:eq}) can be used to find $v$.

To calculate the volumes of the 3-cycles we must identify the spheres in terms of the metric coordinates.
The $5d$ $Y^{p,q}$ metric is:
\begin{eqnarray}
 \label{eq:5d}
\dd s^2_{Y^{p,q}} &=& \frac{1-y}{6} \left( \dd \theta^2 + \sin^2  \theta \dd \phi^2 \right) +
    \frac{\dd y^2}{H(y)} + \frac{H(y)}{36} \left( \dd \beta +  \cos \theta \dd \phi \right)^2 +  \nonumber  \\       
   && + \frac{1}{9} \left( \dd {\psi^\prime} - \cos \theta \dd \phi + 
                                y \left( \dd \beta +  \cos \theta \dd \phi \right) \right)^2 ,
\end{eqnarray}
where
\beq
\label{eq:H(y)}
H(y) = \left( 2 \frac{a-3y^2+2y^3}{1-y} \right)^{1/2}.
\eeq
In these coordinates one can immediately identify the Reeb vector $\xi$ and the contact form $\eta$ in (\ref{eq:5d}):
\beq
\label{eta}
\eta = \frac{1}{3} \left( \dd {\psi^\prime} - \cos \theta \dd \phi + 
                                y \left( \dd \beta +  \cos \theta \dd \phi \right) \right),
\qquad
\xi = 3 \frac{\partial}{\partial \psi^\prime}
\eeq
and the 2-forms $J$ and $J_{KE}$ can be easily derived using the formulae of the previous section.

The coordinates $\phi$ and $\psi^\prime$ are $2 \pi$-periodic, while the azimuthal coordinates 
$\theta$ and $y$ inhabit the ranges $\theta \in [0, \pi]$ and $y \in [y_1,y_2]$,
where the constants $y_1$ and $y_2$ are the smallest two roots of the numerator in (\ref{eq:H(y)}) and are determined by:
\beq
y_{1,2} = \frac{1}{4 p} \left( 2 p \mp 3 q - \sqrt{4p^2-3q^2} \right).
\eeq
These relations also fix the constant $a$ in (\ref{eq:H(y)}). 
The third $2 \pi$-periodic angular coordinate is\footnote{
This identification differs from the one appearing in the literature, see \cite{Ypq}, where
the $2 \pi \ell$-periodic coordinate is claimed to be only the last term in (\ref{alphaNEW:eq}).
We refer the reader to the original paper \cite{EK2}, where the question is discussed in more detail.}:
\beq
\label{alphaNEW:eq}
\tau = \frac{p+q}{2}(\phi+\psi^\prime) - \frac{1}{6 \ell} ( \beta + \psi^\prime) 
\qquad  \textrm{where} \qquad \ell \equiv \frac{q}{3 q^2 -2p^2+p\sqrt{4p^2-3q^2}}.
\eeq 
In \cite{EK2} the gauge invariant variables built from the K\"ahler quotient $\mathbf{C}^4$ coordinates
$z_i$ were matched with the independent non-singular holomorphic functions on $C(Y^{p,q})$. The comparison 
yielded an explicit dependence of the $z_i$'s on the metric coordinates.  
This dependence, of course, included
a free complex parameter. The absolute value of this parameter is the normalization parameter $\Lambda$
used in (\ref{eq:uuvvzzzz}) and
the phase $\lambda$ corresponds to the $U(1)$ gauge of the K\"ahler quotient mentioned in (\ref{U(1):eq}).

With the connection between $z_i$'s and the azimuthal coordinates $\theta$ and $y$ we can identify the 
$z_i=0$ divisors in terms of the metric coordinates. It appears that the bases of the divisors $z_1=0$
and $z_2=0$ correspond to $\theta=0$ and $\theta=\pi$ respectively. Similarly $z_3=0$ and $z_4=0$
are related to $y=y_1$ and $y=y_2$. On the other hand, our three-sphere (defined in \cite{EK2} by $w_2=0$) is given by
the embedding  $\psi^\prime=\textrm{const}$ and $\theta=\theta(y)$, where the latter is a very complicated
function that can be found only numerically. The explicit form of the function, however, is not significant if we
only want to compute the flux of the RR $3$-form through the 3-sphere. 
To this end it is sufficient to know only the boundary conditions which are \cite{EK2}:
\beq
\label{BC:eq}
\theta(y_1)=\pi \qquad \textrm{and} \qquad \theta(y_2)=0.
\eeq
The $3$-form is also a generator of the third cohomology 
class, so the computation provides a decisive check of our $S^3$ identification. 
The RR $3$-form $F_3$ is a real part of the self-dual $(2,1)$ form $G_3$ found 
in \cite{HEK} (see also \cite{EKK}). The RR $2$-form potential is given by:
\beq
C_2 = \frac{p^2-q^2}{16 \pi^2} \left( \frac{1}{1-y} \dd \psi^\prime \wedge \dd \beta + 
                                        \frac{\cos \theta}{1-y} \dd \psi^\prime \wedge \dd \phi
 + \frac{y  \cos \theta}{1-y} \dd \beta \wedge \dd \phi \right), 
\eeq
where $\beta$ is related to the $2\pi$-periodic $\tau$ by (\ref{alphaNEW:eq}).
Substituting $q=1$, $\dd \psi^\prime=0$ and $\theta=\theta(y)$ into $\dd C_{2}$ one can easily verify that the flux is one
as expected\footnote{The formulae:
\beq
\frac{1}{6 \ell} \left( 1 -\frac{1}{y_1} \right) = \frac{p+q}{2}, \qquad
\textrm{and} \qquad
\frac{1}{6 \ell} \left( 1 -\frac{1}{y_2} \right) = -\frac{p-q}{2}
\eeq
are useful for this calculation.} for a representative of the homology class $1\in\Z$.

Again, here only the boundary values (\ref{BC:eq}) of $\theta(y)$ play an important r\^ole.
This is because $C_2$ is globally well-defined except on the submanifolds $y=(y_1,y_2)$ and $\theta=(0,\pi)$,
where the Dirac strings are located (see \cite{EK2}).

\begin{figure}[t]
 \begin{center}
 \includegraphics{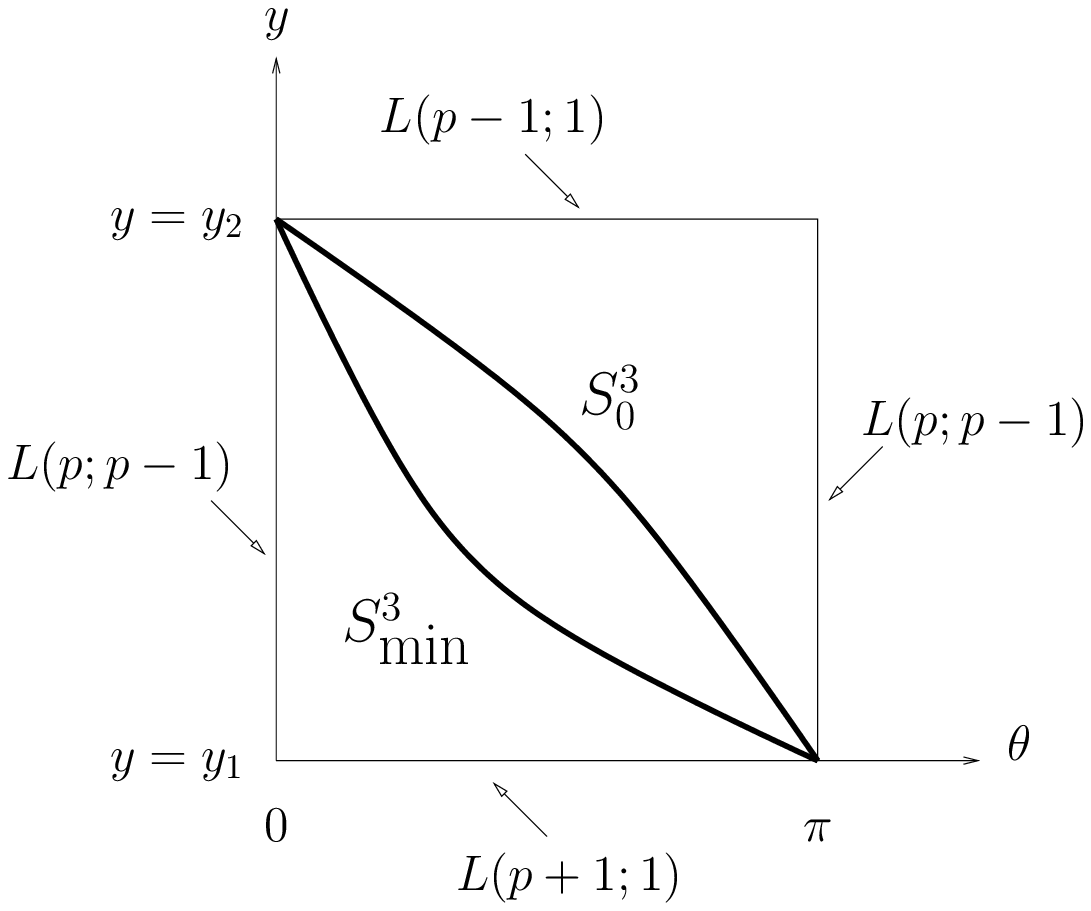}
 \end{center}
 \label{Pict} 
\caption{ This picture shows various $3$-cycles of $Y^{p,1}$ on the $(\theta,y)$ plane. 
Only the lines on the boundary represent supersymmetric cycles.  
For $y=y_{1,2}$ and $\theta=0,\pi$ one finds three different lens spaces (see footnote \ref{Lens}) with the homology classes $(-p)$, $(p-1)$ and $(p+1)$.
One solid curve is the non-BPS three-sphere $S_0^3$ found in \cite{EK2} and 
the other shows the non-BPS $3$-sphere profile $S^3_{\textrm{min}}$ that 
generally minimizes the volume.Note that both curves start and end at the same points.
Also on both curves the $2 \pi$-periodic angles along the $3$-sphere are $\phi$ and $\tau$,
while $\psi^\prime$ is kept constant.}
 \end{figure}

Our strategy, therefore, will be as follows. 
Since for the initial values (\ref{BC:eq}) of the function $\theta=\theta(y)$ and with $\psi^\prime=\textrm{const}$
the homology class of the $3$-cycle is always one,
we may find a function $\theta_{\textrm{min}}=\theta(y)$, which satisfies (\ref{BC:eq})   
and at the same time minimizes the volume of the $3$-sphere. Although we have not found a proof that this ansatz indeed 
leads to the minimal possible volume of the homology class one cycle, this approach is certainly sufficient for
our needs, since our main goal is to show that the volume of the non-BPS cycle 
is smaller than that of a BPS cycle representing the same homology class.  
The situation is summarized in Figure \ref{Pict}.

Finding the profile $\theta(y)_{\textrm{\small{min}}}$, which provides the minimum volume $3$-sphere, amounts to solving
a complicated $2nd$ order differential equation (ODE) with the initial conditions (\ref{BC:eq}).
This equation is provided in the Appendix. We solved it numerically for $p=2,3,4$ and $5$. We then
used this numerical solution to compute the volumes.
Since we are obliged to exploit a numerical approach both for the solution of the ODE and for
the integration of the volume, the final result will be inevitably a bit imprecise.

In order to reach a decisive conclusion regarding the volume comparison, we
will also compute volumes for the following \emph{test} profile:
\beq
\cos (\theta_{\textrm{Test}} (y)) = \frac{2y-y_2-y_1}{y_2-y_1}.
\eeq
The volume calculation for $\theta_{\textrm{Test}}$ turns out to be very accurate.
We will see that even for this probe function the final volumes are usually smaller than their BPS counterparts, though
this is not a solution of the ODE. 
In what follows we report our results for the aforementioned values of $p$.
As in the previous section we will denote the $3d$ bases of the $4d$ $z_i=0$ divisors by $\Sigma_i$.
These BPS $3$-cycles are lens spaces and represent homology classes $-p$, $-p$, $p-1$ and $p+1$
for $\Sigma_1^{(-p)}$, $\Sigma_2^{(-p)}$, $\Sigma_3^{(p-1)}$ and $\Sigma_4^{(p+1)}$ respectively,
where for the sake of clarity we have added superscripts indicating their homology classes.
The volumes of the cycles are:
\begin{eqnarray}
\textrm{Vol} \left( \Sigma_{1,2}^{(-p)} \right) &=& \frac{4 \pi^2}{3} \ell                  \nonumber \\
\textrm{Vol} \left( \Sigma_{3}^{(p-1)} \right)  &=& - \frac{8 \pi^2}{3} \ell y_1 (1-y_1)    \nonumber \\ 
\textrm{Vol} \left( \Sigma_{4}^{(p+1)} \right)  &=& \frac{8 \pi^2}{3} \ell y_2 (1-y_2).  
\end{eqnarray}
Notice that for any $p$ the BPS cycles $\Sigma_{1,2}^{(-p)} \bigcup \Sigma_3^{(p-1)}$ and 
$\Sigma_{1,2}^{(-p)} \bigcup \Sigma_4^{(p+1)}$
have homology classes $\pm 1$. 
For our purposes we will have to compare the cycle with the smallest volume among the two with the non-BPS
homology class one cycles we have described above. 
Finally, the cycles corresponding to $\theta_{\textrm{Test}} (y)$ and $\theta_{\textrm{min}} (y)$
will be denoted by $\Sigma_{\textrm{Test}}$ and $\Sigma_{\textrm{min}}$.

\subsection{$Y^{3,1}$}

For $p=3$ we found:
\beq
\frac{\textrm{Vol} \left( \Sigma_{\textrm{Test}} \right)}{4 \pi^2} \approx 0.133358(0)
\qquad \textrm{and} \qquad  
\frac{\textrm{Vol} \left( \Sigma_{\textrm{min}} \right)}{4 \pi^2} \approx 0.129(8).
\eeq
On the other hand:
\begin{eqnarray}
\frac{\textrm{Vol} \left( \Sigma_{1,2}^{(-3)} \right)}{4 \pi^2} &=& \frac{5 + \sqrt{33}}{144} \approx 0.074615(0)  \nonumber\\
\frac{\textrm{Vol} \left( \Sigma_{3}^{(2)} \right)}{4 \pi^2} &=& \frac{19 + 4\sqrt{33}}{432} \approx 0.083874(3)  \nonumber\\
\frac{\textrm{Vol} \left( \Sigma_{4}^{(4)} \right)}{4 \pi^2} &=& \frac{7 + \sqrt{33}}{216} \approx 0.059002(6).
\end{eqnarray}
We see that although the BPS cycles $\Sigma_i$ have very small volumes compared to $\Sigma_{\textrm{Test}}$
and $\Sigma_{\textrm{min}}$, still the minimal volume of a BPS homology class one cycle is bigger than 
$\textrm{Vol} \left( \Sigma_{\textrm{Test}} \right)$:
\beq
\frac{\textrm{Vol} \left( \Sigma_1^{(-3)} \bigcup \Sigma_4^{(4)} \right)}{4 \pi^2} \approx 0.133617(6)
 > \frac{\textrm{Vol} \left( \Sigma_{\textrm{Test}} \right)}{4 \pi^2} >
 \frac{\textrm{Vol} \left( \Sigma_{\textrm{min}} \right)}{4 \pi^2} 
.
\eeq
As was advertised in the previous section we find that the $3$-sphere with a minimal volume in the homology class $1\in\Z$ 
is non-BPS.

\subsection{$Y^{4,1}$}

For $p=4$ the volumes are:
\beq
\frac{\textrm{Vol} \left( \Sigma_{\textrm{Test}} \right)}{4 \pi^2} \approx 0.100597(5)
\qquad \textrm{and} \qquad  
\frac{\textrm{Vol} \left( \Sigma_{\textrm{min}} \right)}{4 \pi^2} \approx 0.099(1)
\eeq
and:
\begin{eqnarray}
\frac{\textrm{Vol} \left( \Sigma_{1,2}^{(-4)} \right)}{4 \pi^2} &=& \frac{29 + 4\sqrt{61}}{1080} \approx 0.055778(7)  \nonumber\\
\frac{\textrm{Vol} \left( \Sigma_{3}^{(3)} \right)}{4 \pi^2} &=& \frac{43 + 5\sqrt{61}}{1728} \approx 0.047483(3)  \nonumber\\
\frac{\textrm{Vol} \left( \Sigma_{4}^{(5)} \right)}{4 \pi^2} &=& \frac{91  + 11 \sqrt{61}}{2880} \approx 0.061428(0).
\end{eqnarray}
Thus for the homology class one cycles we have:
\beq
\frac{\textrm{Vol} \left( \Sigma_1^{(-4)} \bigcup \Sigma_3^{(3)} \right)}{4 \pi^2} \approx 0.103262(1)
 > \frac{\textrm{Vol} \left( \Sigma_{\textrm{Test}} \right)}{4 \pi^2} >
 \frac{\textrm{Vol} \left( \Sigma_{\textrm{min}} \right)}{4 \pi^2} 
.
\eeq
So, again, we can draw very firm conclusions.

\subsection{$Y^{5,1}$}

For $p=5$ the volumes are:
\beq
\frac{\textrm{Vol} \left( \Sigma_{\textrm{Test}} \right)}{4 \pi^2} \approx 0.080690(4)
\qquad \textrm{and} \qquad  
\frac{\textrm{Vol} \left( \Sigma_{\textrm{min}} \right)}{4 \pi^2} \approx 0.080(0)
\eeq
and
\begin{eqnarray}
\frac{\textrm{Vol} \left( \Sigma_{1,2}^{(-5)} \right)}{4 \pi^2} &=& \frac{47 + 5\sqrt{97}}{2160} \approx 0.044557(5)  \nonumber\\
\frac{\textrm{Vol} \left( \Sigma_{3}^{(4)} \right)}{4 \pi^2} &=& \frac{133 + 13\sqrt{97}}{5400} \approx 0.048339(8)  \nonumber\\
\frac{\textrm{Vol} \left( \Sigma_{4}^{(6)} \right)}{4 \pi^2} &=& \frac{73  + 7 \sqrt{97}}{3600} \approx 0.039428(3)
\end{eqnarray}
Thus for the homology class one cycles we have:
\beq
\frac{\textrm{Vol} \left( \Sigma_1^{(-5)} \bigcup \Sigma_4^{(6)} \right)}{4 \pi^2} \approx 0.083985(9)
 > \frac{\textrm{Vol} \left( \Sigma_{\textrm{Test}} \right)}{4 \pi^2} >
 \frac{\textrm{Vol} \left( \Sigma_{\textrm{min}} \right)}{4 \pi^2}.
\eeq
This time the BPS cycle has $5\%$ bigger volume the non-BPS cycle
and the results suggest that the ratio grows as one increases $p$.

\subsection{$Y^{2,1}$}

The $p=2$ case is very special, since the cycle $\Sigma_{3}^{(1)}$
is both BPS and homology class one.
Its volume is:
\beq
\frac{\textrm{Vol} \left( \Sigma_{3}^{(1)} \right)}{4 \pi^2} = \frac{31 + 7\sqrt{13}}{432} \approx 0.130182(5).
\eeq 
We also found that:
\beq
\frac{\textrm{Vol} \left( \Sigma_{\textrm{min}} \right)}{4 \pi^2} \approx 0.185(0).
\eeq
Although we were not able to calculate $\textrm{Vol} \left( \Sigma_{\textrm{min}} \right)$
with higher accuracy, the answer is significantly bigger than $\Vol \left( \Sigma_{3}^{(1)} \right)$
and so for $Y^{2,1}$ the minimum volume $3$-sphere is apparently BPS.
It is natural to propose that a BPS cycle will minimize the volume in the homology class $1$ for all of the $Y^{p,p-1}$
spaces, including $Y^{1,0}=T^{1,1}$, since in these cases $\Sigma_3^{(1)}$ is both BPS and also a homology class $1\in\Z$ cycle.  This is a natural proposal, as in these cases the divisor $C(\Sigma^(1)_3)$, which is a cone over a representative of the element $1$ in the third homology group, is an irreducible variety.

\section{Energies of BPS Cycles} \label{townsec}

BPS states are necessarily either stable or marginally stable.  In particular, no state with the same conserved 
charges may have a lower energy.  In the present context, this implies for example that D3-branes wrapping 
generalized calibrated cycles in a Sasaki-Einstein 5-fold cannot decay.  If there were two such cycles representing 
the same homology class, then a transition between them would be allowed and so such branes must have the same energy.
We have already seen examples of generalized calibrated 3-cycles in the same homology class with different volumes, 
therefore the Born-Infeld contribution to their energies, which measures their volumes, must be precisely compensated 
by the Wess-Zumino contribution to their energy.  Such a cancellation is guaranteed by the general arguments of 
\cite{gcalpapers1}.  In this section we will apply these general arguments to the specific case of generalized 
calibrated 3-cycles on a Sasaki-Einstein 5-fold.  As the RR gauge 
potential $C_4$ extends along 4 Sasaki-Einstein directions, and the BPS D3-branes only extend along 3, 
one may have concluded that the Wess-Zumino terms play no role.  We will now see that this is not the case.

For simplicity, we will consider a D3-brane with a vanishing gauge potential wrapped on the 3-cycle $\Sigma^3$.  
Then the DBI action just produces its volume.  If the cycle is calibrated by a calibration $\alpha_3=J_{KE}\wedge\eta$, 
then the volume is just equal to the integral of $\alpha_3$.  Now we want to show that the total energy of this D-brane 
is the same as that of any other D-brane wrapped on any other calibrated 3-cycle $\widetilde{\Sigma^3}$ in the same homology class as $\Sigma^3$. 
Formally this is equivalent to showing that the energy of a brane on $\Sigma^3$ plus an anti-brane (with no absolute value in the DBI energy) on $\widetilde{\Sigma^3}$ is equal to zero, 
which is in turn equivalent to showing that the energy of a brane on a calibrated cycle of trivial homology is equal to zero.  
It is this last statement that we will show.  So we may assume that $\Sigma^3$ is homologically trivial, and so there exists 
some 4-chain $B^4$ whose boundary is $\Sigma^3$.

How does one calculate the energy associated with the DBI and Wess-Zumino terms in the action?  First of all, energy is the charge corresponding to some translational symmetry.  Therefore one must choose a direction in which to perform the translation.  The BPS condition imposes that the energies with respect to a vector $\xi$ given by the preserved SUSYs (\ref{xi}) is the same for all BPS cycles.  Therefore we will be interested in the energy with respect to this vector.  We claim that if all of the fields and connections have a zero Lie derivative with respect to $\xi$, then the energy density with respect to $\xi$ is just the interior product $i_\xi$ of the Lagrangian density.  To calculate the total energy, one pulls back the energy density to a surface at a constant time and then integrates.  

This prescription is perhaps more familiar in electrostatics.  
The Lagrangian of a particle of charge one contains the 1-form potential $A_\mu$.  
The energy of the particle is just the interior product of $A_\mu$ with the time vector ${\partial_t}$, 
which is:
\beq
i_{\partial_t}A=A_0
\eeq
often called the scalar potential.  The scalar potential is only well defined up to an additive constant, 
which cancels when one considers the difference between the energies of two particles.  We are interested 
in the difference in the energies between two branes, and so this additive constant will cancel.  
Notice that gauge transformations can change $A_0$ by more than just a constant, but when the magnetic field 
is time-independent one may always choose a gauge in which $\dot{A_i}=0$ and so $A_0$ is the energy of the electron.

In our case the field strength $F_5$ is time-independent and so the Wess-Zumino energy is just $C_4$ 
contracted with our temporal vector.  On the other hand the DBI energy is just the \mbox{4-volume} contracted with 
the temporal vector.  The $\xi$ part of the temporal vector contracted with the 4-volume form on the 
D3-brane worldvolume gives a form with one leg along time, which vanishes when pulled back to a spatial slice, 
therefore only the $i_{\partial_t}(\dd t \wedge \dd \Vol_{\Sigma^3})$ contributes to the DBI energy\footnote{
Here $\Sigma^3$ is a 3-cycle wrapped by the D3-brane.}, 
which for a generalized calibrated cycle is just the calibration form $\alpha_3$.  
Summarizing, the Wess-Zumino energy density is $i_\xi C_4$ and the DBI energy density is the spatial volume form.

In a supersymmetric configuration, the metric and the field strengths are invariant under a translation along the 
Reeb vector field $\xi$.  Therefore there exists a gauge such that the RR gauge potential $C_4$ is also invariant:
\beq
\mathcal{L}_{\xi}C_4 = \dd i_\xi C_4 + i_\xi F_5 = 0.
\label{lie}
\eeq
In this gauge the Wess-Zumino contribution to the energy density of our brane is just $i_\xi C_4$.  
Technically, one needs to use the sum of the Reeb vector of the Sasaki-Einstein manifold with that of the $AdS_5$, 
but the latter will not contribute to the energy for a D3-brane which is only 1-dimensional in the AdS directions, like ours. 
 The Wess-Zumino energy can then be calculated using (\ref{lie}) and Stokes' theorem:
\beq
\mathcal{E}_{\rm{WZ}}=\int_{\Sigma^3} i_\xi C_4=\int_{B^4} \dd i_\xi C_4=-\int_{B^4} i_\xi F_5 
       = -\int_{B^4} \dd \alpha_3  = -\int_{\Sigma^3} \alpha_3,
\eeq
where in the fourth equality we used the property (\ref{ig}) of generalized calibrations.  
We have just argued that the integral of $\alpha_3$ is the DBI energy, and so we have shown that the 
Wess-Zumino energy over a trivial calibrated cycle is precisely minus the Wess-Zumino energy, and so the total energy 
is equal to zero.  Therefore the energies of branes on homologous generalized calibrated cycles are equal. In other words, 
homologous BPS D-branes have the same energy.

The above argument is well-known.  In the Sasaki-Einstein case we have considered in this paper, we may be 
a bit more explicit.  
We saw in Section \ref{calsec} that the RR field strength is:
\beq
F_5= 2 J_{KE}\wedge J_{KE}\wedge \eta
\eeq
and so its interior product with respect to the Reeb vector is just:
\beq
i_\xi F_5= 2 J_{KE}\wedge J_{KE} = \dd \alpha_3.
\eeq
The explicit expression for $\xi$ and $\eta$ in the $Y^{p,q}$ case appear in (\ref{eta}) and for any Sasaki-Einstein 5-manifold $\dd \eta = 2 J_{KE}$.
It is also not too difficult to find the 4-chain $B^4$ for $Y^{p,q}$'s. 
Let us consider 3-cycles $\Sigma_1^{(-p)}$ and $\Sigma_3^{(p-q)}$ introduced in Section \ref{Stansec}.
Obviously, $\Sigma^3=(p-q) \Sigma_1^{(-p)} + p \Sigma_3^{(p-q)}$ is a trivial 3-cycle.
The cone $C(B^4)$ over the 4-chain $\partial  B^4 = \Sigma^3$ is then given by $\Arg (z_1^{p-q} z_3^p) = 0$,
where $z_i$'s are the K\"ahler quotient coordinates of Section \ref{Stansec}.
In particular, recall that $\Sigma_i$ is a 3-dimensional base of the cone $z_i=0$. 
The variables $z_1$ and $z_3$ have quotient charges $-p$ and $p-q$ respectively and so this product is gauge invariant.
In the dual gauge theory it corresponds to a mesonic operator.


\section{Prospects: Energies of Non-BPS Cycles}

The energy of a non-BPS cycle is greater than that of a BPS cycle in the same homology class.  
The difference is the failure of the bound (\ref{ineq}) to be saturated.  In other words, the difference in 
energies between a BPS and a non-BPS cycle in the same homology class, including both DBI and Wess-Zumino contributions, is:
\beq
\Delta \mathcal{E}= \Vol_{\Sigma} - \int_{\Sigma} \alpha_3
 \label{nbpse},
\eeq
where, again, $\Sigma$ is the cycle wrapped by the D3-brane.
The generalized calibration condition (\ref{ineq}) guarantees that this difference is never negative, and so 
BPS cycles minimize energy.  For example, in the case of $Y^{3,1}$ a brane wrapping the cycle $\Sigma_{\textrm{min}}$ 
will have an energy which is greater than that of a BPS cycle in the same homology class ($1\in\Z=\H_3(Y^{3,1})$) by:
\beq
\frac{\Delta \mathcal{E}}{4 \pi^2} \approx 0.056(7)
\eeq
despite the fact that all such BPS cycles have greater volumes.  In other words, the flux causes a D3-brane wrapped on 
$\Sigma_{\textrm{min}}$ to expand.

It would be interesting to interpret the values of the energies of these operators in the dual gauge theory. 
Gubser and Klebanov \cite{GubserKlebanov} have argued that, in the case of BPS operators on $T^{1,1}$, the volume of the 
cycle corresponds to the conformal weight.  This conjecture has subsequently been extended to BPS cycles in other
Sasaki-Einstein's.  If one may find the 3-cycle dual to a given non-chiral operator with a baryonic charge, 
even in $T^{1,1}$, then (\ref{nbpse}) may be used to compute the energy of that operator and thus to try 
to determine the corresponding gauge theory quantity.

\section*{Acknowledgements}

We would like to thanks R. Argurio, A. Hanany, A. Uranga  and especially J. Sparks for invaluable discussions and 
correspondences.
It is also  a great pleasure to thank \mbox{D.~N.~E. Persson}.

S.~K. is supported in part by the Belgian Federal Science Policy Office
through the Interuniversity Attraction Pole P6/11, in part by the European
Commission FP6 RTN programme MRTN-CT-2004-005104 and in part by the
``FWO-Vlaanderen'' through project G.0428.06.

\section*{Appendix}

Here we report the differential equation that we had to solve in order to
find the three-sphere with the minimal volume.
To derive this equation one has to substitute the $\theta=\theta(y)$, $\dd \psi=0$ ansatz into (\ref{eq:5d})
and to calculate the $3$-cycle volume from the induced metric. The variation with respect to $\theta(y)$
then gives the following equation:
\begin{eqnarray}
& & \left( \left( \frac{w(y)  \left( \frac{1-y}{6} \sin(\theta)^2 + \frac{v(y)}{9} \cos(\theta)^2 \right) }
                     {  \frac{1-y}{6} {\theta^\prime_y}^2 + \frac{1}{w(y) v(y)} } \right)^{1/2}
                 \frac{1-y}{6} \theta^\prime_y  \right)^\prime_y   =                                            \\
 & & \quad  \quad
    =\left( \frac{w(y) \left( \frac{1-y}{6} {\theta^\prime_y}^2 + \frac{1}{w(y) v(y)} \right)}
                     {\frac{1-y}{6} \sin(\theta)^2 + \frac{v(y)}{9} \cos(\theta)^2 }   \right)^{1/2}   
           \left(  \frac{1-y}{6} -  \frac{v(y)}{9} \right) \sin(\theta)  \cos(\theta),         
   \nonumber       
\end{eqnarray}
where 
\beq
 w(y) = \frac{2(a-y^2)}{1-y}
\qquad \textrm{and} \qquad
 v(y) = \frac{a-3y^2+2 y^3}{a-y^2}.
\eeq

\bibliographystyle{unsrt}

\begin{thebibliography}{--} 



\bibitem{Harvey:1982xk}
R. Harvey and H.~B. Lawson Jr.,
{\it Calibrated geometries},
Acta Math.\, {\bf 148},  47, (1982)
  



\bibitem{gcalpapers1}
J.~Gutowski and G.~Papadopoulos,
  {\it AdS calibrations}, [arXiv:hep-th/9902034]   

J.~Gutowski, G.~Papadopoulos and P.~K.~Townsend, 
  {\it Supersymmetry and generalized calibrations}, 
  [arXiv:hep-th/9905156]

J.~Gutowski,
  {\it Generalized calibrations}, 
  [arXiv:hep-th/9909096]

P.~K.~Townsend, {\it PhreMology: Calibrating M-branes}, [arXiv:hep-th/9911154]



\bibitem{gcalpapers2}

L.~Martucci and P.~Smyth,
  {\it Supersymmetric D-branes and calibrations on general \mbox{N = 1} backgrounds},
  [arXiv:hep-th/0507099] 

J.~Evslin and L.~Martucci,
  {\it D-brane networks in flux vacua, generalized cycles and calibrations},
  [arXiv:hep-th/0703129]


D. Martelli and J. Sparks, 
{\it G-Structures, Fluxes and Calibrations in M-Theory},
[{\tt arXiv:hep-th/0306225}]



  

\bibitem{HackettJones:2004yi}

  E.~J.~Hackett-Jones and D.~J.~Smith,
{\it Type IIB Killing spinors and calibrations},
  [arXiv:hep-th/0405098]


  
  
  
  
\bibitem{Ypq}



J.~P. Gauntlett, D.~Martelli, J. Sparks and D. Waldram, 
{\it {S}asaki-{E}instein metrics on ${S}^2 \times {S}^3$}, 
[{\tt arXiv:hep-th/0403002}]  



D.~Martelli and J.~Sparks, 
{\it Toric geometry, {S}asaki-{E}instein manifolds and a new infinite class of {AdS/CFT} duals}, 
[{\tt arXiv:hep-th/0411238}]  



M.~Bertolini, F.~Bigazzi and A.~L. Cotrone, {\it New checks and subtleties for
  ads/cft and a-maximization},  
[{\tt arXiv:hep-th/0411249}]



S. Benvenuti, S. Franco, A. Hanany, D. Martelli and J. Sparks, 
{\it An infinite family of superconformal quiver gauge theories with
  {S}asaki-{E}instein duals}, 
[{\tt arXiv:hep-th/0411264}] 
  
  



  
\bibitem{contact}


  D.~Martelli, J.~Sparks and S.~T.~Yau,
  {\it Sasaki-Einstein manifolds and volume minimisation},
  [arXiv:hep-th/0603021]
 
  C.~B\"ar, {\it Real Killng spinors and Holonomy},
  Commun.\ Math.\ Phys.\  {\bf 154} (1993) 509-521


  
  
  
  
  


\bibitem{Labc}


M. Cvetic, H. Lu, D.~N. Page and C.~N. Pope, {\it New {E}instein-{S}asaki
  spaces in five and higher dimensions}, 
[{\tt  arXiv:hep-th/0504225}]  



D. Martelli and J. Sparks, 
{\it Toric {S}asaki-{E}instein metrics on ${S}^2 \times {S}^3$},  
[{\tt arXiv:hep-th/0505027}]  



S. Benvenuti and M. Kruczenski, {\it From {S}asaki-{E}instein spaces to quivers
  via {BPS} geodesics: ${L}^{p,q,r}$},
[{\tt arXiv:hep-th/0505206}] 



S. Franco, A. Hanany, D. Martelli, J. Sparks, D. Vegh and B. Wecht, 
{\it Gauge theories from toric geometry and brane
  tilings}, 
[{\tt arXiv:hep-th/0505211}] 


A. Butti, D. Forcella and A. Zaffaroni, 
{\it The dual superconformal theory for ${L}^{p,q,r}$ ma\-ni\-folds},  
[{\tt arXiv:hep-th/0505220}]  
 

M. Cvetic, H. Lu, D.~N. Page and C.~N. Pope, 
{\it New {E}instein-{S}asaki and {E}instein spaces from {K}err-de {S}itter},
[{\tt arXiv:hep-th/0505223}] 


S. Kuperstein, O. Mintkevich and J. Sonnenschein,
{\it On the pp-wave limit and the BMN structure of new Sasaki-Einstein spaces},
[{\tt arXiv:hep-th/0609194}] 

  
  

\bibitem{Sparks}

 J.~Sparks, private communication.


 

 

\bibitem{WittenBranesBaryonsAds}
E. Witten,
{\it Baryons and branes in anti-de Sitter space},
[{\tt arXiv:hep-th/9805112}]



 
\bibitem{GubserKlebanov}
S.~S. Gubser and I.~R. Klebanov,
{\it Baryon spectra and AdS/CFT correspondence},
[{\tt hep-th/9808075}]










\bibitem{HBKandB}


D. Berenstein, C.~P. Herzog and I.~R. Klebanov,
{\it Baryon spectra and AdS/CFT correspondence},
[{\tt hep-th/0202150}]


C.~E. Beasley,
{\it BPS branes from baryons},
[{\tt hep-th/0207125}]




\bibitem{KW}

I.~R. Klebanov and E. Witten,
{\it Superconformal Field Theory on Threebranes at a Calabi-Yau Singularity},
[{\tt arXiv:hep-th/9807080}]





\bibitem{CdO}

P. Candelas and X.~C. de la Ossa,
{\it Comments on Conifolds},
Nucl.\ Phys.\ B {\bf 342}, 246-268 (1990).




\bibitem{EK1}
J. Evslin and S. Kuperstein, 
{\it Trivializing and orbifolding the conifold's base},
[{\tt arXiv:hep-th/0702041}]




\bibitem{EK2}
J. Evslin and S. Kuperstein, 
{\it Trivializing a Family of Sasaki-Einstein Spaces},
[{\tt arXiv:0803.3241}]






\bibitem{HEK}
C.~P. Herzog, Q.~J. Ejaz and I.~R. Klebanov,
  {\it Cascading RG flows from new Sasaki-Einstein manifolds},
[{\tt arXiv:hep-th/0412193}]



\bibitem{EKK}
J. Evslin, C. Krishnan and S. Kuperstein, 
{\it Cascading quivers from decaying D-branes},
[{\tt arXiv:0704.3484}]



 
\end{thebibliography}

\end{document}

\bibitem{Intriligator:2003jj}
K. Intriligator and B. Wecht,
{\it The exact superconformal R-symmetry maximizes a},
[{\tt hep-th/0304128}]

\end{document}